\documentclass[preprint,prX]{revtex4}

\usepackage[utf8x]{inputenc}
\usepackage{graphicx}
\usepackage{amsmath}
\usepackage[version-1-compatibility]{siunitx}
\usepackage{chemfig}
\usepackage{color}
\usepackage{braket}
\usepackage{todonotes}

\begin{document}

\title{Fermi orbital self-interaction corrected electronic structure of molecules beyond local density approximation}
\author{T. Hahn}
\email{torsten.hahn@physik.tu-freiberg.de}
\author{S. Liebing}
\author{J. Kortus}
\affiliation{Institute of Theoretical Physics, TU Bergakademie Freiberg, 09596 Freiberg, Germany}
\author{Mark R. Pederson}
\affiliation{Department of Chemistry, Johns Hopkins University, Baltimore, Maryland 21218, USA}

\keywords{DFT, SIC}


\begin{abstract}
 The correction of the self-interaction error (SIE) that is inherent to all standard density functional theory (DFT) calculations is an object of increasing interest. In this article we apply the very recently developed Fermi-orbital based approach for the self-interaction correction (FOSIC) \cite{pederson_communication:_2014,Pederson2015} to a set of different molecular systems.
Our study covers systems ranging from simple diatomic to large organic molecules. We focus our analysis on the direct estimation of the ionization potential from orbital eigenvalues.

\end{abstract}

\maketitle

\section{Introduction}

Electronic structure calculations within the DFT framework have become an invaluable tool for physicists, chemists and material scientists due to low computational costs and sufficient accuracy of this method \cite{Cohen2012}. Within DFT, the quality of the results and therefore of subsequent derived properties is strictly bound to the functional used to evaluate the exchange and correlation energy ($E_{XC}$). The estimation of $E_{XC}$ is the only approximation that is needed in the expression of the total energy within DFT \cite{Martin2004}.

The search for improved descriptions of this contribution has lead from the simple LDA and GGA approaches to various levels of hybrid functionals for example B3LYP, HSE and meta hybrid GGA's like M06 \cite{B3LYP,HSE,MS06}.

Some of the failures of the different functionals have been considered to be intrinsic to the DFT approach. Examples are the dissociation energies of two center three electron systems \cite{gruning2001failure}, overestimation of the magnetic coupling \cite{kortus2001hamiltonian,park2004properties,pederson_density-functional_1985}, description of charge-transfer systems \cite{dreuw2003long,baruah2009dft} and ionization potentials \cite{Ruiz1996,Kummel2008}. In nearly all cases these faults can be related to the self-interaction error which is, of course to a different degree, inherent in all functionals used in todays DFT implementations. This effect arises from the spurious interaction of an electron with itself. In the Hartee-Fock framework this contribution is totally cancelled by the exchange contribution and this is also the main reason why the hybrid functionals often predict properties closer to experiment \cite{ruiz1997toward}. The self-interaction error is partially cancelled by the fraction of HF density that is used in the construction of $E_{XC}$.


The self-interaction problem within the DFT approach has been identified very early, already in the process the method was developed. Perdew and Zunger (PZ) proposed a method for self-interaction correction (SIC) \cite{Perdew1981} back in 1981. 
The major drawback of the so called PZ-SIC method is, that one needs to evaluate an orbital dependent exchange correlation potential. This results in excessive computational cost and therefore in practice PZ-SIC is only applied to small model systems or simplified Hamiltonians \cite{svane1988anti,Johnson1994,Klupfel2012}.
Some other approaches were developed and the focus was always to somehow avoid the calculation of the orbital dependent $E_{XC}$ \cite{Ciofini2003,Tsuneda2014}.


The aim of this paper is to investigate the performance of the FOSIC approach in the field of molecular systems. We focus on the evaluation of the vertical ionization potentials (IPs) for test cases, ranging from very small to large conjugated molecular systems. Although the first IPs could be directly calculated using the DFT extension of the Koopmans' theorem \cite{Koopmans1934104,PhysRevLett.49.1691} the negative of the HOMO energy is typically several electron volts too small with respect to the experimental values.
This effect, due to orbital self-interaction error, varies with the physical extent of the occupied and unoccupied orbitals and can be responsible for non-systematic errors that cause unphysical charge-transfer between chemically disparate components in molecular assemblies and solids. With the inclusion of the self-interaction correction the HOMO energies become much closer to the experimental IPs. Due to the fact that this property is directly related to the self-interaction error this provides a benchmark which is essential to validate the FOSIC method. 
It should be noted here that the Koopmans' theorem in Hartree Fock (HF) states that the HF eigenvalue is equal to the difference between the fully relaxed ground-state of the N electron system and the ground-state of the N-1 electron system when the energy of the N-1 electron system is relaxed within the space spanned by the orbitals of the N-electron system. Hence in HF, where the Koopmans' theorem is exact, one should expect that the HF eigenvalue overestimates the fully relaxed HF $\Delta$-SCF ionization energy and, in the weak-correlation limit, the experimental ionization energy as well. A similar variational Koopmans' theorem has been proven for the PZ-SIC theory \cite{pederson_density-functional_1985} but it has been noted that the non-quadratic 
dependence of the exchange-correlation functional removes the algebraically exact differences between the total energy of the N and N-1 electron and requires small, but difficult to account for, deviations between the eigenvalue and the energy difference. These differences have been referred to in the past \cite{heaton1985density} as "non-Koopmans" corrections and again more recently by Dabo et al \cite{PhysRevB.82.115121,dabo2013donor}. In contrast to HF, even in the weak-correlation limit, the analytically dependence of the energy functional on density prevents one from simply stating that the SIC-DFT eigenvalue should always overestimate the experimental ionization potential. However since to lowest order the SIC-DFT eigenvalue does not include relaxation of the system out of the space of orbitals spanned by the N-electron ground state, it may be reasonable to anticipate that SIC-DFT eigenvalues which overestimate experiment are more likely in most cases.
\\

The paper is organized as follows: We will first give a short review on the essential ideas of SIC and the FOSIC approach to DFT. We will only focus on the general concepts that are necessary for this paper. For a detailed review we point the reader to the original papers published by Pederson and coworkers and the references given therein \cite{pederson_communication:_2014,Pederson2015}. Second, we will describe the  details of the used calculation procedure. Finally we will discuss the obtained results and identify crucial points for the further development of the method.

\section{Theoretical background}
 
\subsection{Fermi Orbital SIC} 
\label{sub:fo_sic}
The FOSIC method is based on the original  Perdew and Zunger approach \cite{perdew_self-interaction_1981}.
The DFT functional is expressed in the following way:
\begin{align}\label{eq:xc1}
E^{SIC}_{xc}&=E_{xc}^{LSD}[n_\uparrow,n_\downarrow]-E^{PZ-SIC}_{xc}
\end{align}
with
\begin{align}
 \label{eq:sic1}
E^{PZ-SIC}_{xc} &= -\sum\limits_{\alpha,\sigma} \left\{ U \left[ \rho_{\alpha,\sigma}\right] + E^{approx}_{xc}\left[\rho_{\alpha,\sigma},0\right] \right\}\,.
\end{align}

In equation \ref{eq:sic1} the orbitals $\{\phi_{\alpha\sigma}\}$ define orbital densities according to \(n_{\alpha\sigma} (\mathbf{r})\approx | \phi_{\alpha\sigma}(\mathbf{r})|^2 \). The term $ U\left[ \rho_{\alpha,\sigma}\right]$ is the  exact self-Coulomb energy and  $E^{approx}_{xc}\left[\rho_{\alpha,\sigma},0\right]$ is the approximated exchange-correlation energy. Essentially the PZ-SIC method is a correction to the standard $E_{XC}$ given by the local spin density approximation.     
As discussed in \cite{heaton1983self,heaton1987new,pederson1988localized} the best option to solve equation \ref{eq:sic1} is the use of localized, canonical orbitals. Thereby one has to make sure that the Hermitian Lagrange multiplier matrix of the orbitals has to fullfill a O($N^2$) localization equation
\begin{align}
 \label{eq:localize}
\left\{H_{0,\sigma} + V_{i\sigma}^{SIC} \right\}\ket{\phi_{i\sigma}} &= \sum\limits_j\lambda_{ij}^\sigma \ket{\phi_{j\sigma}} \\ \bra{\phi_{i\sigma}} V^{SIC}_{i,\sigma} &- V^{SIC}_{j,\sigma} \ket{\phi_{j\sigma}}=0 \nonumber
\end{align}
with respect to the orbital density $n_{\alpha\sigma}$.


In principle any means for parameterizing a unitary transformation may be used to solve the localization equations. However, it has been recognized that the full optimization of the unitary transformation, at least when used within existing functionals, leads to an expression for the total energy that is not necessarily size extensive . Ensuring size extensivity within the PZ-SIC method is difficult since it is hard to systematically derive approximate functionals that will deliver a negative SIC energy for the highest occupied orbitals of a heavy atom. A new approach, which constrains the functional to be unitarily invariant and fixes the problem associated with size extensivity, was to use localized orbitals on the basis of the Fermi hole the so called Fermi orbitals \cite{luken_localized_1982,luken_localized_1984}.

A set of KS orbitals can be transformed to Fermi orbitals in any point of space according to
\begin{align}
F_{i\sigma} (\mathbf{r}) &= \frac{\rho(\mathbf{a}_{i\sigma},\mathbf{r})}{\sqrt{\rho_\sigma(\mathbf{a}_{i\sigma})}},\\ F_{i\sigma}(\mathbf{r}) &= \frac{\sum_\alpha\psi^*_{\alpha\sigma}(\mathbf{a}_{i\sigma})\psi_{\alpha\sigma}(\mathbf{r})}{\sqrt{\{\sum_\alpha\left|\psi_{\alpha\sigma}(\mathbf{a}_{i\sigma})\right|^2\}}}\equiv \sum_\alpha F^\sigma_{i\alpha} \psi_{\alpha\sigma}(\mathbf{r}).
\end{align}

Due to the procedure proposed in \cite{pederson_communication:_2014} it is possible to bypass the direct solution of the localization equations \ref{eq:localize} and therefore the computational effort needed has the advantage to have the scaling of standard DFT calculations. 
%
The expression for the exact exchange energy using a single Slater determinant is given by
\begin{align}
E^x &= -\frac{1}{2} \sum_\sigma \int d^3\mathbf{r} \int d^3\mathbf{a} \frac{|\sum_\alpha\psi^*_{\alpha\sigma}(\mathbf{r})\psi_{\alpha\sigma}(\mathbf{a})|^2}{|\mathbf{r}-\mathbf{a}|}\,.
\end{align}
According to \cite{pederson_communication:_2014} this can be rewritten to
\begin{align}
E^x_\sigma &=-\frac{1}{2}\int d^3\mathbf{r} \rho_\sigma (\mathbf{r}) \int d^3\mathbf{a} \left\{\frac{\rho_\sigma(\mathbf{r},\mathbf{a})\rho_\sigma(\mathbf{a},\mathbf{r})}{\sqrt{\rho_\sigma(\mathbf{r})}\sqrt{\rho_\sigma(\mathbf{r})}}\right\}\frac{1}{|\mathbf{r}-\mathbf{a}|}
\end{align}
where the term 
\begin{equation}
 \frac{\rho_\sigma(\mathbf{r},\mathbf{a})\rho_\sigma(\mathbf{a},\mathbf{r})}{\sqrt{\rho_\sigma(\mathbf{r})}\sqrt{\rho_\sigma(\mathbf{r})}} = |F(\mathbf{a})|^2
\end{equation}
is the exchange hole and therefore equivalent to the square of the Fermi orbital.
The constructed Fermi orbitals depend parametrically on a set of $N_\sigma$ positions which correspond to quasi-classical electronic positions. By the availability of gradients of the Fermi orbitals \cite{Pederson2015} one has a way to minimize the self interaction corrected total energy as function of the positions of the Fermi orbitals.


\subsection{Obtaining the Fermi orbital positions} 
\label{sub:fo_positions}

An important challenge of the FOSIC method is that one needs a set of initial Fermi orbitals in the beginning of the calculation. For atoms and very small molecules this could be obtained by a brute-force method (see \cite{pederson_communication:_2014}) or by using `chemical intuition'. Clearly this may become not feasible for large molecules or automated calculations.

\begin{figure*}[tb]
 \includegraphics[width=0.99\textwidth]{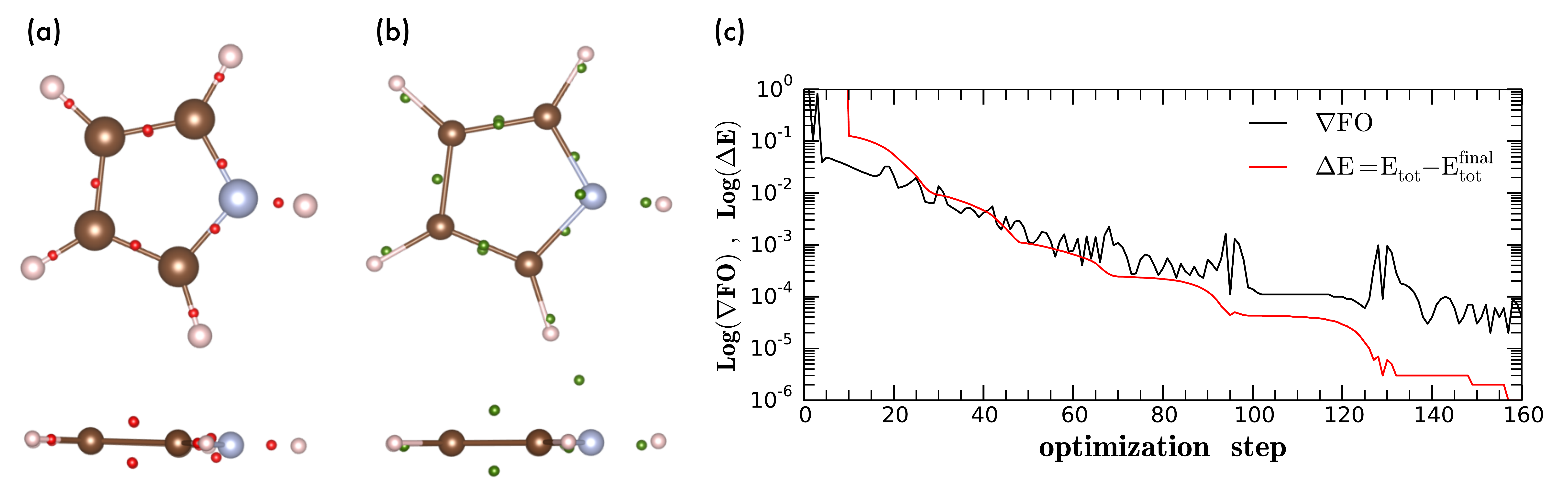} 
 \caption{Example of the optimization process of the FO positions: a) The red markers show the initial positions of the FO centroids as obtained from the Wannier analysis. b) The green markers show the final converged positions of the FO centroids. c) Development of the total energy and the forces on the FO's as a function of optimization step. A version of the \textsc{FIRE} algorithm \cite{Bitzek2006} was used.}
 \label{fig:FO_optimize}
\end{figure*}

As a straightforward way to obtain approximate initial positions for the Fermi orbitals we propose using maximally localized Wannier orbitals (MLWF) \cite{PhysRevB.56.12847}. It has been claimed that Wannier orbitals and Fermi orbitals are identical under certain conditions \cite{PSIK-Pederson} and they seem therefore as a natural choice to develop an initial-guess. Here we followed the approach of \cite{PhysRevLett.94.026405} and considered a set of $N_{occ}$ eigenstates of $\{ \ket{u_m} \}$. The total energy is invariant with respect to unitary transformations $U_{mn}$ among the eigenstates
\begin{align} \label{eq:mlwf1}
 \ket{w_n} = \sum^{N_{occ}}_{m=1} U_{mn}\ket{u_m}\; .
\end{align}
The unitary matrix $U$ is chosen such that the resulting $N_{occ}$ orbitals $\{w_n(r)\}$ minimize their total quadratic spread, given by
\begin{align} \label{eq:mlwf2}
 S_n^2 = \sum_n \left( \braket{w_n|r^2|w_n} - \braket{w_n|\mathbf{r}|w_n}^2 \right) = \sum_n \left( \braket{r^2}_n - \mathbf{\bar{r}}^2_n \right) \; .
\end{align}
Each MLWF is characterized by a value of its quadratic spread, $S_n^2$  and its centre $\mathbf{\bar{r}}_n$.\\
We carried out the following procedure to obtain the initial guess for the Fermi orbital positions: 
\begin{enumerate}
 \item perform a standard DFT calculation of the molecule of interest
 \item construct a set of MLWF's according to equations \ref{eq:mlwf1} and \ref{eq:mlwf2}
 \item take the centers  $\mathbf{\bar{r}}_n$ as initial positions for the subsequent FOSIC calculation
\end{enumerate}
These initial positions then need to be optimized in order to minimize $E_{tot}$ of the system of interest. In our investigation it turned out, that the  $\mathbf{\bar{r}}_n$ is a quite robust initial guess. In Fig. \ref{fig:FO_optimize}a) and b) we show for one example the initial guess and the final Fermi orbital positions after the optimization procedure. The quality of the guess is in general reasonable and reflects the qualitative details of the final Fermi orbital positions. However, we observed that with the current localization scheme the resulting Wannier centers appear to be too close to the nearest neighbour atoms. This is not a principle problem since the subsequent optimization of these initial positions always leads to correct placement of the final Fermi orbitals (see Fig. \ref{fig:FO_optimize}). A drawback is that in the current implementation the optimization algorithm needs a large number of steps to find the final positions of the Fermi orbitals.

\subsection{Computational details}

We have used a modified version of the \textsc{nrlmol} all-electron DFT code \cite{jackson_accurate_1990,pederson_strategies_2000} that uses large Gaussian-orbital basis sets \cite{porezag_optimization_1999} for the representation of the electron wavefunctions. The standard DFT potentials as well as the FOSIC potential is calculated on a mesh using the variational mesh technique \cite{pederson_variational_1990}. The PW92 functional \cite{Perdew1992} within the local density approximation (LDA) was used for the standard LDA as well as for the FOSIC calculations.

The GGA calculations utilizing the B3LYP functional for exchange and correlation \cite{schafer_fully_1992,weigend_balanced_2005} were carried out using version 2.9 of the \textsc{ORCA} DFT package \cite{neese_configuration_2001}. We utilized the def2-TZVP basis set \cite{Weigend2005} in order to avoid artifacts related to small basis sets.

The geometry of all molecules have been optimized either with LDA/PW92 or the GGA/B3LYP functional to obtain relaxed structures with atomic forces below \SI{0.01}{eV/\angstrom}.

\subsection{Ionization potential}

The primary ionization potentials for LDA and B3LYP were taken as the negative of the eigenvalue of the highest occupied molecular orbital $\mathrm{IP=-\epsilon_{HOMO}}$ (see Table \ref{table:mainresults1} and \ref{table:mainresults2}). We further used the LDA/PW92 result as our starting point for the FOSIC calculations. Following the description in sec. \ref{sub:fo_positions} we constructed a set of maximally localized Wannier orbitals from the Kohn-Sham orbitals of the LDA/PW92 calculation. The centers of the Wannier orbitals were used as initial positions for Fermi orbitals. A subsequent optimization of the Fermi orbital positions in order to minimize the total energy was carried out by using a conjugated gradient method. For this task the forces on the Fermi orbitals were calculated. The optimization algorithm was stopped once these forces on electrons were below  $10^{-5}$ Hartree/bohr and  $\mathrm{-\epsilon_{HOMO}}$  was taken from the resulting electronic configuration.

The $\Delta\mathrm{SCF}$ values are obtained by taking the difference of the total energies between the neutral and the positive charged molecules without relaxing the geometry of the charged state.

\section{Results and discussion}

We have chosen two different groups of molecules in order to test how the FOSIC method performs within structures of different complexity. Test set 1 (see table \ref{table:mainresults1}) includes very small di- and triatomic molecules and small to medium sized molecules with double and triple bonds covering most of the main group elements. The 2nd test set consists of different organic aromatic structures ranging from simple 5 or 6 member rings with different substitutes to conjugated structures with increasing size (from two up to five condensed rings).\\

\begin{table}[b]
  \caption{Test set 1: Experimental values for the experimental ionization potential (IP) and the respective calculated negative of the HOMO energy for different methods (all values in [eV]). MAE is the mean average error. As discussed in the text, an exact Koopmans' theorem should overestimate the ionization energy. It is only exact in the limit of weak correlation, a quadratic expression for the exchange-correlation energy, and if electronic relaxation, outside the space spanned by the N-electron space of orbitals, is unimportant.}
  \label{table:mainresults1}
 \begin{tabular}{cccccc}
 \hline
 molecule & exp. IP & $-\epsilon_{\mathrm{H}}^{LDA}$ & $-\epsilon_{\mathrm{H}}^{FOSIC}$ & $-\epsilon_{\mathrm{H}}^{B3LYP}$ & IP$^{LDA}_{\Delta_{SCF}}$ \\
 \hline
$\mathrm{N_2}$             & 15,56 & 10,43 & 18,21  & 11,8 & 15,65\\
$\mathrm{O_2}$             & 12,07 &  6,92 &  16,06  & 7,01 & 12,74\\
$\mathrm{H_2S}$            & 10,46 & 6,36 & 11,69  & 7,13 & 11.01\\

$\mathrm{CO}$              & 14,10 & 9,12 & 15,75  & 10.40 & 14.10\\
$\mathrm{CO_2}$            & 13,77 & 9.32 & 16,26  & 10.32 & 14.04\\
$\mathrm{H_2O}$            & 12,62 & 7,33 & 15,42  & 8.51  & 13.11\\
$\mathrm{HCN}$             & 13,60 & 9.18 & 15,41  & 10.07 & 14.11\\
$\mathrm{LiCl}$            & 10,01 & 6.00 & 11,52  & 6.76  & 10.33\\
$\mathrm{LiH}$             &  7,9  & 4.39 & 9.16   & 5.28  & 8.19 \\
~ \\
MAE                         &  ~   & 4.6  & 2.2 & 3.6 & 0.4 \\

 \hline
 \end{tabular}
\end{table}

\begin{table*}
  \caption{Test set 2: Experimental values for the experimental ionization potential (IP) and the respective calculated negative of the HOMO energy for different methods (all values in [eV]). MAE is the mean average error.}
  \label{table:mainresults2}
 \begin{tabular}{ccccccc}
 \hline
 molecule & exp. IP & $-\epsilon_{\mathrm{H}}^{LDA}$ & $-\epsilon_{\mathrm{H}}^{FOSIC}$ & $-\epsilon_{\mathrm{H}}^{ADSIC}$ & $-\epsilon_{\mathrm{H}}^{B3LYP}$ & IP$^{LDA}_{\Delta_{SCF}}$ \\
 \hline
$\mathrm{ethan}$           & 11,52 & 8.05 & 14,74  & -    & 9.31  & 11.74\\
$\mathrm{ethen}$           & 10,51 & 6.97 & 12,63  & 12.1 & 7.56  & 11.02\\
$\mathrm{ethen-flourene}$  & 10,12 & 6.37 & 13.55  & -    & 7.36  & 10.25\\
$\mathrm{ethin}$           & 11,40 & 7.36 & 12,94  & -    & 8.09  & 11.77\\
$\mathrm{benzen}$           & 9,24 & 6,55 & 9,41   & 9.96   & 6.97 & 9.55 \\
$\mathrm{fulvene}$          & 8,55 & 5,56 & 9,01   & 8.91 & 6.07 & 8.60 \\
$\mathrm{furan}$            & 8,89 & 5,85 & 9,56   & 9.85 & 6.39 & 9.19 \\
$\mathrm{pyrazine}$         & 9,4  & 5,93 & 10,67  & 9.36 & 7.00 & 9.15 \\
$\mathrm{pyridine}$         & 9,6  & 5,97 & 10,17  & 9.39 & 7.07 & 9.48 \\
$\mathrm{pyrimidine}$       & 9,73 & 6,07 & 10,79  & 9.52 & 7.11 & 9.39 \\
$\mathrm{thiophene}$         & 8,87 & 6,06 & 9,68   & 9.75 & 6.56 & 9.19 \\
$\mathrm{dimethylsilaethene}$& 8,3 & 5,06 & 9,64   & 8.6 & 5.62 & 8.19 \\
$\mathrm{TCNQ}$             & 7,29 & 7,24 & 11,80  & - & 7.56 & 9.32 \\
$\mathrm{naphthalene}$      & 8,14 & 5.69 & 8,73   & - & 6.03 & 8.16\\
$\mathrm{anthracene}$       & 7,44 & 5,19 & 8,03   & - & 5.44 & 7.33\\
$\mathrm{chrysene}$         & 7,60 & 5.48 & 8.27   & - & 5.73 & 7.45\\
$\mathrm{picene}$           & 7,51 & 5.47 & 8.15*  & - & 5.71 & 7.28\\
~ \\
MAE                         &  ~   &  2.9 & 1,4 & ~ & 2.3 & 0.3 \\

 \hline
 \end{tabular}
\end{table*}


As expected the uncorrected $-\epsilon_{\mathrm{H}}$ values (no FOSIC) for the test set of very small molecules have a large mean average error (MAE) of $4.6\,\mathrm{eV}$. The experimental ionization potential is generally underestimated my several electron volts. A much smaller error of $0.4\,\mathrm{eV}$ is found for the $\Delta \mathrm{SCF}$ approach. The values do agree very well with the experimental data due to the fact that for $\Delta \mathrm{SCF}$ the self-interaction error cancels at least partially. In some examples like $\mathrm{CO}$ the $\Delta \mathrm{SCF}$-IP matches the experimental value almost perfectly. 
The  $-\epsilon_{\mathrm{H}}$ values obtained by the FOSIC approach on the other hand offer IP values with a mean average error of $2.2\,\mathrm{eV}$. That is a clear improvement compared to the uncorrected LDA values. The $-\epsilon_{\mathrm{H}}$ values obtained for the popular B3LYP functional have a mean average error of $3.6\,\mathrm{eV}$ which is significant larger than the FOSIC error. The B3LYP IP values are in between the LDA and FOSIC values. Both LDA and B3LYP underestimate the experimental ionization potential values while FOSIC typically overestimates.

\begin{figure}[tb]
 \includegraphics[width=0.9\columnwidth]{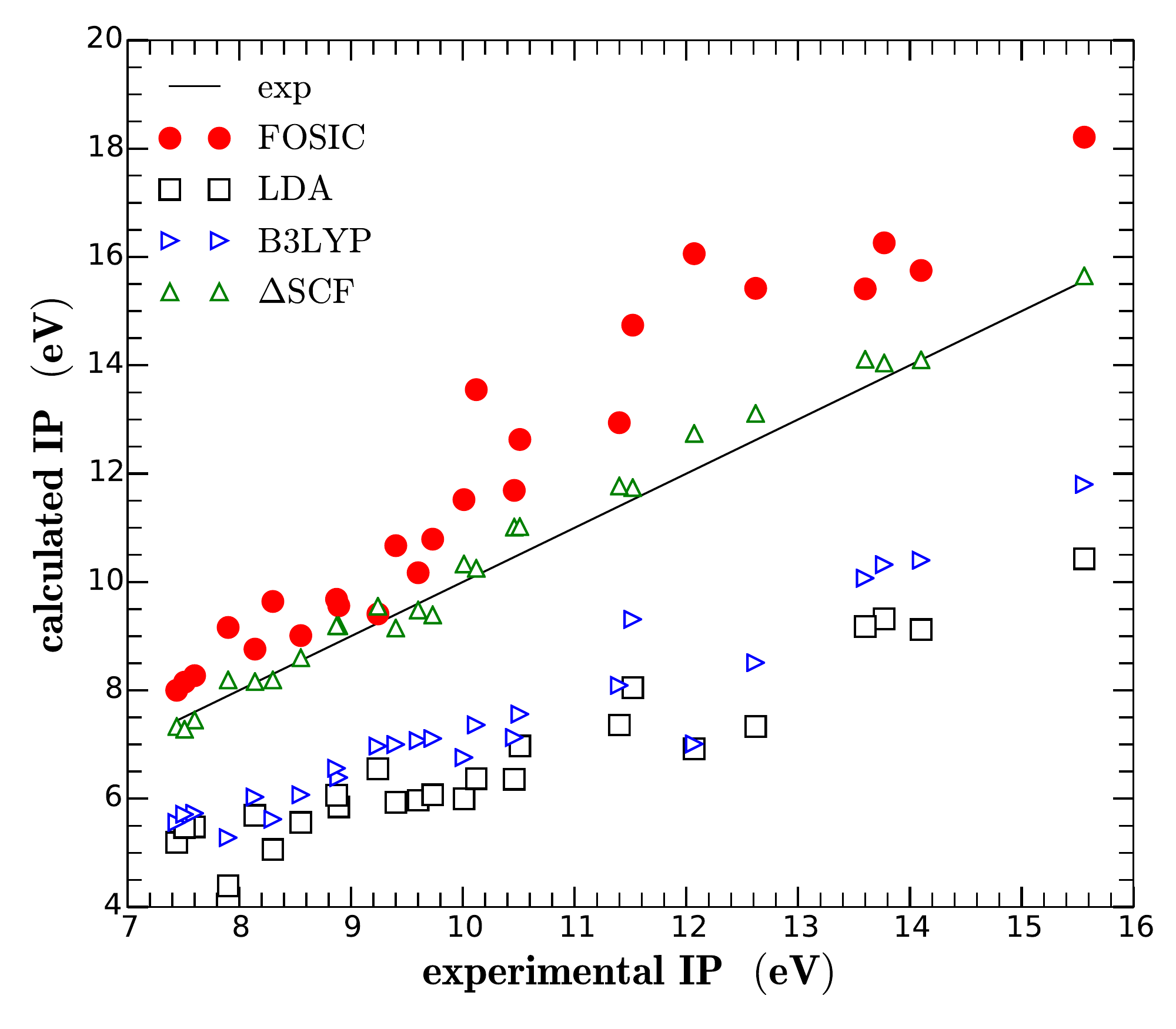} 
 \caption{Comparison of the ionization potentials obtained from orbital eigenvalues and experimental values.}
 \label{fig:IP_values}
\end{figure}

For the medium to large molecules presented in table \ref{table:mainresults2} no clear trend is visible. Overall the obtained $-\epsilon_{\mathrm{H}}$ values for the FOSIC calculations have a lower mean average error of $1.4\,\mathrm{eV}$ compared to $1.9\,\mathrm{eV}$ the standard DFT values. For B3LYP ($MAE=2.3\,\mathrm{eV}$) one notes again that FOSIC does perform better in general. Again we see that FOSIC overestimates the ionization potential. However the magnitude of the overestimation is different. The deviation of the FOSIC IP's from the experimental values does not show a systematic behavior. All the differently calculated IP values, in comparison to there experimental counter part are shown in figure \ref{fig:IP_values}.
However, it is worth noting that especially for the larger molecules the deviation from experiment is rather small compared to LDA or B3LYP. in the case of chrysene and picene FOSIC provides even a qualitative better result. While LDA gives nearly identical IP's for the two molecules FOSIC does reproduce the experimental finding, that picene has a $\approx 0.1\,\mathrm{eV}$ lower IP than chrysene.

We also included some values from the literature obtained by the average density SIC approach \cite{Ciofini2003}. The values agree very well with our results. The average density SIC results are based on the Becke88 GGA functional \cite{PhysRevA.38.3098} which somewhat limits the direct comparison to our PW92 FOSIC values. However it supports the correctness of the FOSIC implementation and makes us optimistic to see further improvement of orbital eigenvalues when applying FOSIC to GGA type functionals. 

\begin{figure}[tb]
 \includegraphics[width=0.9\columnwidth]{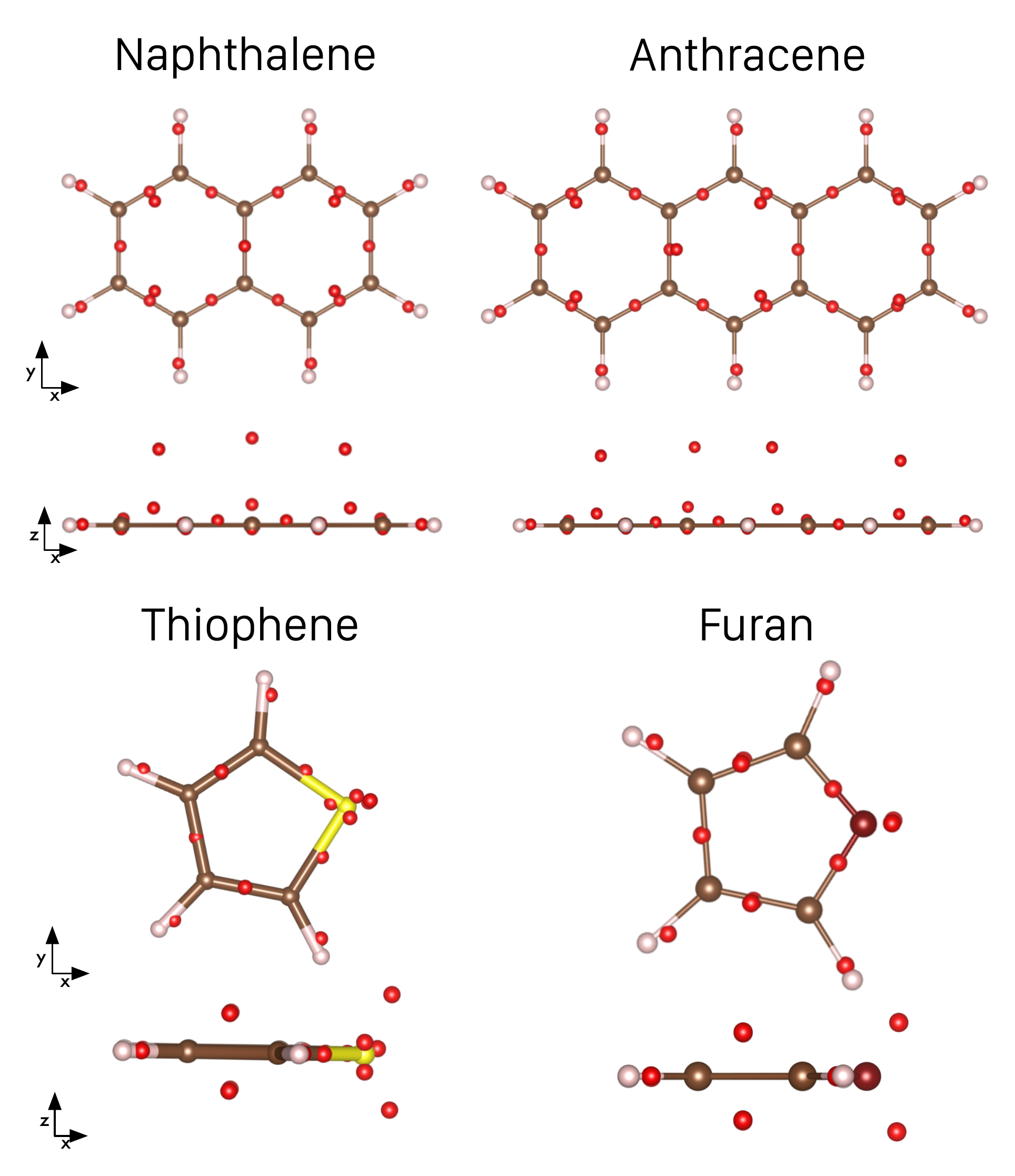} 
 \caption{Optimized positions of the Fermi orbitals for different structures. The red dots represent the final Fermi orbital positions.}
 \label{fig:FO_transfer}
\end{figure}

Further, we found that the positions of the Fermi orbitals are transferable between similar molecules as previously reported \cite{pederson_communication:_2014}. It was for example very successful to use the Fermi orbital positions obtained for benzene as starting values for the 6-rings of the poly-aromatic hydrocarbons naphthalene to picene. The subsequent optimization of such an initial guess converged almost immediately and the resulting $-\epsilon_{\mathrm{H}}$ values are in very good agreement with the experimental data. In Fig. \ref{fig:FO_transfer} we show such a case. For the case of naphthalene and anthracene one sees that the relative positions of the Fermi orbitals with respect to the neighbouring carbon atoms within the aromatic rings are almost identical. 
The Fermi orbitals representing single bonds are always located in the $xy$-plane in the middle between the neighbouring C atoms. On the other hand, the Fermi orbitals representing the double bonds are 0.3 and 1.4~\AA~ above the molecular plane. The same argument holds for the comparison of the structurally similar thiophene and furan molecules (Fig. \ref{fig:FO_transfer} bottom). Again the distribution of the Fermi orbitals inside the 5-ring is nearly identical. The difference between furan and thiophene can be identified in the additional electrons of the sulfur atom, that form a perfect tetrahedron and the distance of the Fermi orbitals forming the free electron pairs (0.90~\AA ~sulfur vs 0.78~\AA ~oxygen). The larger distance in case of thiophene is an indication of the larger Coulomb repulsion from the additional core electrons. The out of plane Fermi-Orbital positions in the systems with six-membered carbon rings have also been reported by Pederson and Baruah in applications to Benzene \cite{pederson_baruah_2015}

\begin{figure}[tb]
 \includegraphics[width=0.9\columnwidth]{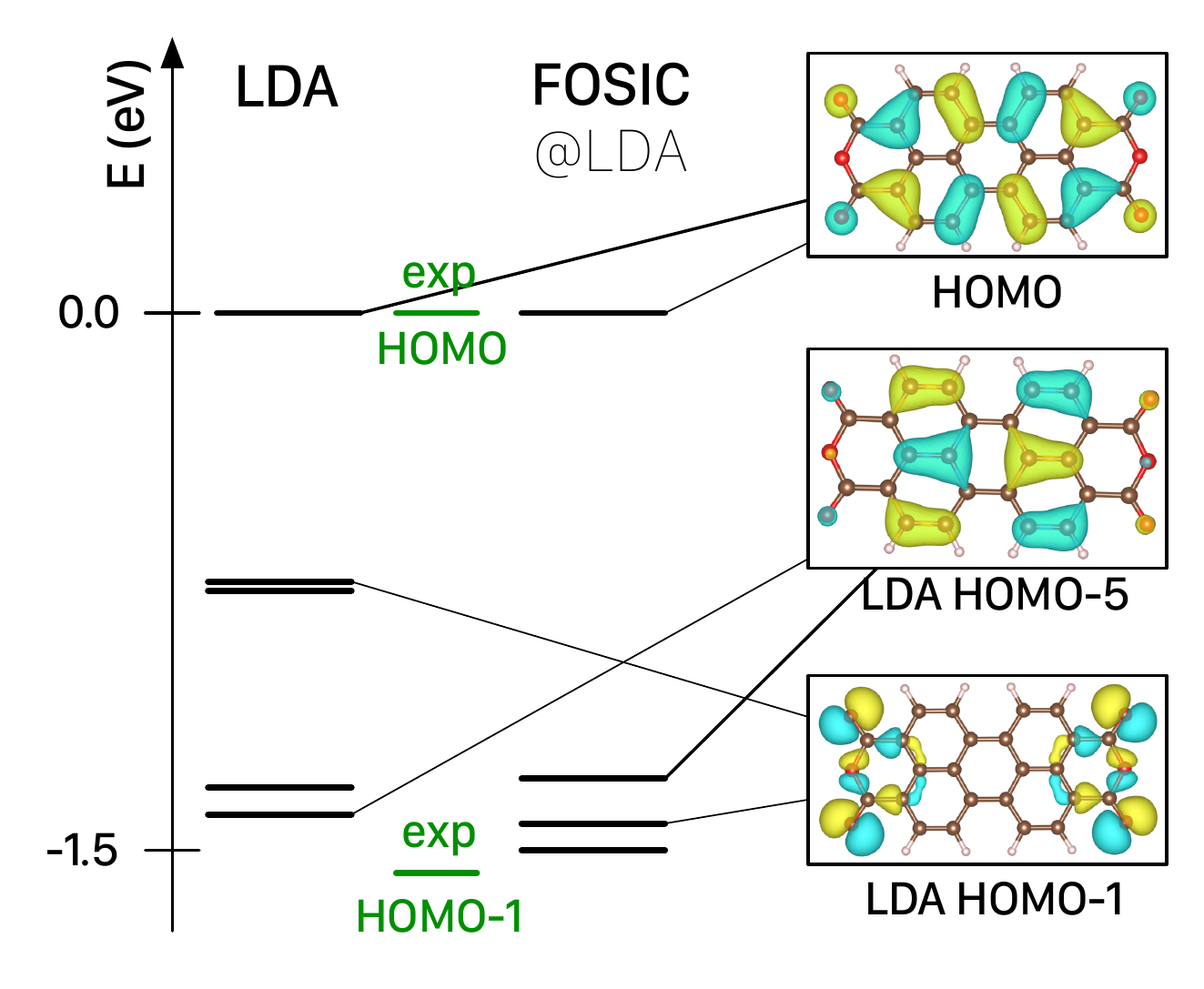} 
 \caption{Position of the orbital eigenvalues of PTCDA from LDA and FOSIC method compared to experimental data (green) obtained by gas phase photoelectron spectroscopy (values extracted from \cite{PhysRevB.73.195208}) together with a view of an isosurface plot of the respective molecular orbitals as obtained by the LDA and FOSIC calculations (HOMO-2 - HOMO-4 not shown for clarity). HOMO levels are aligned for better comparsion.}
 \label{fig:FO_levels}
\end{figure}

In some cases we even obtain a more realistic level ordering from the FOSIC calculation than from LDA. One representative example is given in Fig. \ref{fig:FO_levels}. Here we computed the electronic structure of PTCDA \footnote{3,4,9,10-Perylenetetracarboxylic dianhydride}, a molecule that consists of an aromatic core and oxygen end groups whose structure is also shown in Fig. \ref{fig:FO_levels}. If one compares the calculated to experimental values, obtained by gas phase photoemission spectroscopy, one clearly sees that LDA offers not at all an appropriate description of the measured values. Hence the orbital energies calculated by FOSIC are in much better agreement. The orbital analysis shows that within LDA the orbitals delocalized over the central aromatic ring system form the HOMO and HOMO-5 and they agree reasonable well with experiment. The LDA HOMO-1 to HOMO-4 orbitals however are localized at the oxygen end groups yielding LDA energies that are significant too high due the self-interaction error and therefore they appear wrongly in between HOMO and HOMO-5. With FOSIC the energy of these orbitals is shifted down below the delocalized HOMO-5 resulting in a level ordering in very good agreement with the experiment. The delocalized orbitals do suffer much less from the spurious self-interaction because the contributions to the Coulomb-potential come from a much larger distance. In general one has to state that in cases where delocalized states interact substantially with localized ones FOSIC delivers a qualitatively correct picture while standard LDA (and also GGA) probably does not. 

\section{Summary and outlook}

We have presented a validation of the FOSIC approach used in DFT. We have tested the performance of the approach on the direct estimation of the ionization potential from the orbital eigenvalues. 
We obtain ionization potentials that are comparable to the $\Delta \mathrm{SCF}$ results and therefore in very good agreement with experiments for all molecular systems. We have shown that FOSIC offers a significant improvement over standard LDA and B3LYP calculated ionization potentials. 

The investigations suggests that FOSIC, in its current state, is capable of handling organic molecules in a straightforward way. However there is still room for improvement. A crucial point is the improvement of the initial guess for the FO positions. One possibility may be using the Edmiston and Ruedenberg criteria \cite{Edmiston1963}, which uses the Coulomb self-interaction instead of the quadratic spread for the construction of the maximally localized Wannier orbitals. This is however far more demanding from an implementations point of view \cite{Subotnik2004} and is therefore left for further studies. The observation of transferable positions in structurally similar molecules appears to be very promising for application of the method to large molecules.
The current implementation is based on LSDA, therefore an extension to other functionals in particular to GGA ones is highly desirable.  A longer term goal might be to design functionals that are explicitly corrected for self-interaction error using a systematic unitarily invariant FOSIC methodology discussed here.

\begin{acknowledgements}
S.L. wants to thank the Friedrich-Ebert-Stiftung for the financial support.
Further we would like to thank the ZIH Dresden for providing computational resources. M.R.P  thanks the Texas Advanced Computing Center (TACC) and National Energy Research Scientific Computing (NERSC) for computer time to test the minimization methods.
\end{acknowledgements}


\bibliographystyle{h-physrev.bst}
\bibliography{sic}

\end{document}